# A 128-point Multi-Path SC FFT Architecture

Shun-Che Hsu & Shen-Jui Huang
*Intelligo Technology Inc.*
Hsinchu City, Taiwan
shenray.ee95g@g2.nctu.edu.tw

Sau-Gee Chen & Shin-Che Lin
*Insitute of Electronics*
*Natioanl Chiao Tung University*
Hsinchu City, Taiwan
sgchen@g2.nctu.edu.tw

Mario Garrido
*Department of Electrical Engineering*
*Universidad Politécnica de Madrid*
Madrid, Spain
mario.garrido@upm.es

*Abstract*—This paper presents a new radix-$2^k$ multi-path FFT architecture, named MSC FFT, which is based on a single-path radix-2 serial commutator (SC) FFT architecture. The proposed multi-path architecture has a very high hardware utilization that results in a small chip area, while providing high throughput. In addition, the adoption of radix-$2^k$ FFT algorithms allows for simplifying the rotators even further. It is achieved by optimizing the structure of the processing element (PE). The implemented architecture is a 128-point 4-parallel multi-path SC FFT using 90 *nm* process. Its area and power consumption at 250 MHz are only 0.167 mm$^2$ and 14.81 mW, respectively. Compared with existing works, the proposed design reduces significantly the chip area and the power consumption, while providing high throughput.

*Keywords—SC FFT, radix-$2^k$ FFT, multi-path structure*

## I. INTRODUCTION

Fast Fourier transform (FFT) is widely used in advanced communication systems, particularly in those that adopt the transmission techniques of orthogonal frequency division multiplexing (OFDM) and single-carrier frequency division multiple access (SC-FDMA). For instance, the most advanced mobile R15 5G NR system, IEEE 802.11ax and 802.11ay. These systems demand a throughput an order of magnitude higher than their predecessors, which leads to the need of high-performance FFTs, specifically the parallel FFT architectures.

Since last decades, many pipelined FFT hardware architectures have been proposed: Single-path delay feedback (SDF) [1], which processes serial data and includes feedback loops; multi-path delay feedback (MDF) [2-10], which processes parallel data and includes feedback loops; single-path delay commutator (SDC) [11-12], which processes serial data without feedback loops; multi-path delay commutator (MDC) [13-15], which processes parallel data without feedback loops; and the recently proposed single-path serial commutator (SC) FFT architecture [16], which makes use of circuits for bit-dimension permutation of serial data. The SC FFT architecture has the advantage of eliminating the low utilization problem of SDF FFTs thanks to a novel data management scheme that reorders butterfly and rotation operations. As a result, the number of adders and multipliers in the processing element (PE) of each stage are reduced by half.

Although the SC FFT is very hardware-efficient, it considers only a single serial input and output data streams.

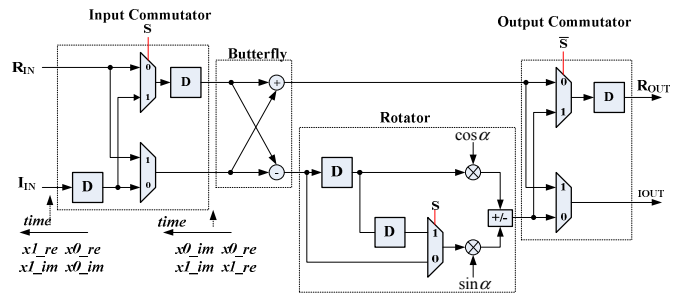

Fig. 1. Processing element of the SC FFT.

Consequently, it is hard to meet the throughout requirement of those advanced communication systems mentioned above. In this paper, we propose a new parallel multi-path FFT architecture by integrating parallel copies of SC FFT architecture and thoroughly optimizing the multi-path FFT architecture by fully utilizing all kinds of optimization techniques as much as possible. As a result, all the merits of high hardware utilization, high area efficiency, low power consumption and high throughput can be simultaneously achieved for the proposed MSC FFT architecture As a demonstration of the proposed design concept, we design a 128-point 4-parallel MSC FFT architecture that explores and exploits high-radix FFT algorithms, i.e., radix-$2^3$ and $2^4$ FFT algorithm for obtaining an optimized result as much as possible. Moreover, we incorporate the rotator allocation techniques in [17], in order to increase the hardware utilization even further.

This paper is organized as follows. In Section II, we describe the background of this work. In Section III, the proposed MSC FFT architecture is presented. In Section IV, experimental results and comparison with previous works are provided, followed by Section V, the conclusion.

## II. BACKGROUND

### A. The fast Fourier transform

The *N*-point discrete Fourier transform (DFT) for an input sequence, $x[n]$, is defined as

$$X[k] = \sum_{n=0}^{N-1} x[n] W_N^{kn}, \qquad k = 0, 1, \ldots, N\text{-}1, \qquad (1)$$

where $X[k]$ is the *k*th DFT coefficient and $W_N^{nk} = e^{-j2\pi nk/N}$. Fast Fourier transform (FFT) algorithms are applied to reduce the computational cost of the DFT, leading to a complexity of $O(N\log_2 N)$ which is much less than $O(N^2)$ for direct computation of the DFT. During the FFT process, the major basic computation involved is the multiplication of a data with so called twiddle factors $W_L^m = e^{-j2\pi m/L}$, where $m$ and $L$ are some positive integers. The operation is also equivalent to a vector rotation by an angle of $2\pi m/L$.

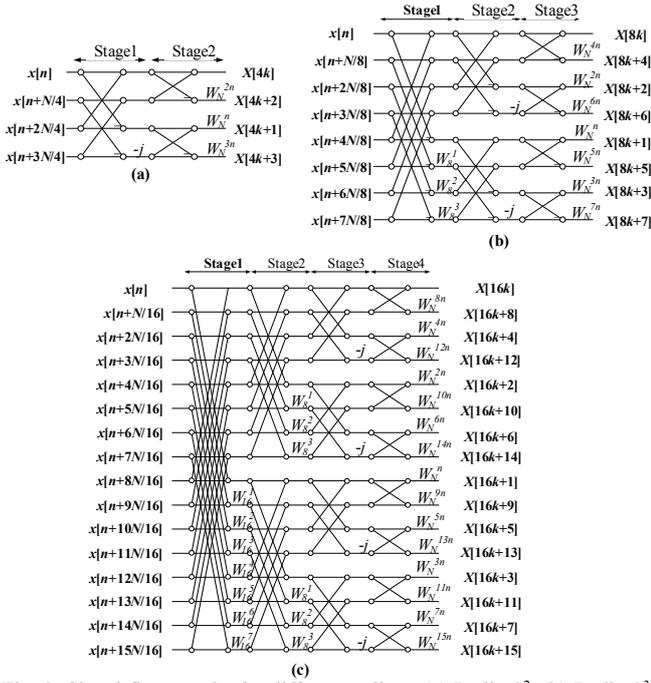

Fig. 2. Signal flow graphs for different radices. (a) Radix-$2^2$. (b) Radix-$2^3$. (c) Radix-$2^4$.

*B. The SC FFT architecture*

Fig. 1 shows the processing element of the SC FFT in [16]. It consists of a half-butterfly, a half-rotator, input and output circuits for data reordering. Note that data arrives serially at a rate of one sample per clock cycle, being the two parallel paths in the figure for the real and imaginary parts of the same sample.

*C. Symmetric angle sets*

A *symmetric angle set* (SAS) [17,18] is a set of angles of the form $n\pi/2 \pm \alpha$, where $n = 0,\ldots,3$ and $\alpha \in [0, \pi/4]$. Any rotation in a symmetric angle set can be calculated as a rotation by an angle $\alpha \in [0, \pi/4]$, a trivial rotation and an exchange of the real and imaginary part of the rotation coefficient.

*D. M-rotator*

An *M*-rotator or *M*-rot [18] is a rotator that can rotate a number of angles in *M* different symmetric angle sets. For instance, a rotator that rotates by 0º, 45º and 135º is a 2-rot as it rotates angles in the symmetric angle sets $n\pi/2$ and $n\pi/2 \pm \pi/4$. Likewise, the twiddle factor $W_8$ is a 2-rot, the twiddle factor $W_{16}$ is a 3-rot and the twiddle factor $W_{32}$ is a 5-rot. In general, a twiddle factor $W_L$ is an $L/8+1$-rot. This means that there are $L/8+1$ angles in the range $[0, \pi/4]$ and the rest of the angles can be obtained by utilizing the symmetry property.

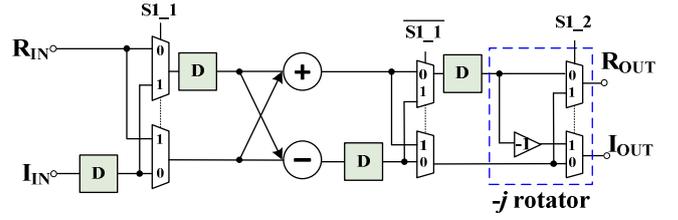

Fig. 3. PE with $W_4$ rotator (PE_W4).

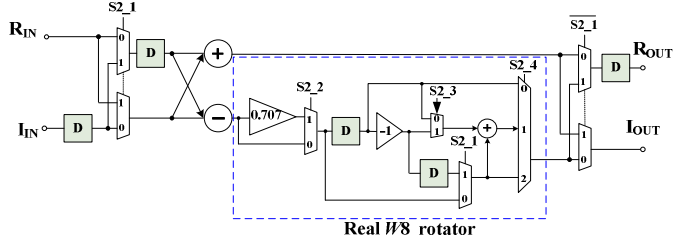

Fig. 4. PE with $W_8$ rotator (PE_W8).

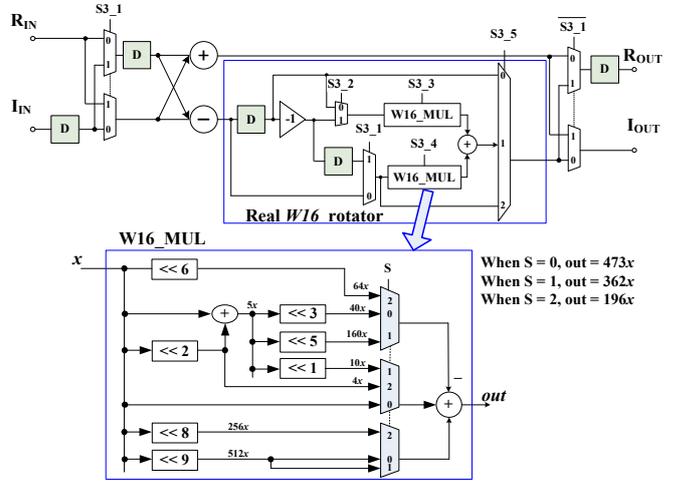

Fig. 5. PE with $W_{16}$ (PE_W16).

III. PROPOSED MULTI-PATH SC FFT ARCHITECTURE

*A. Simplification of the PE for radix-$2^k$*

Fig. 2 shows the flow graphs of radices $2^2$, $2^3$ and $2^4$ FFTs which are based on decimation-in-frequency (DIF) decomposition. The proposed architecture is based on these building blocks. One can observe that the rotation operations in the flow graph only appear at the lower edges of the butterfly outputs. This translates into the fact that only the lowest path of the PE in SC FFT needs to be rotated. As a result, the PEs in Figs. 3 to 6 are derived.

Fig. 3 shows the realization of the PE for the stages where a $W_4$ rotation is applied. In this case, only a $-j$ rotation is needed. The S1_2 signal selects a rotation by 1 or $-j$. Fig. 4 shows the realization of the PE for the stages where $W_8$ rotation is operated. In this case, the constant 0.707 in $W_8^1 = 0.707 - 0.707j$ and $W_8^3 = -0.707 - 0.707j$ is moved to the input of the rotator. As a result, the rotator only requires one constant multiplier. With proper control signal setting, this PE calculates all the rotations only with $W_8$ twiddle factor. The constant multiplication by 0.707 can be implemented as four simple shift-and-add operations for 12-bit precision, as can be easily shown.

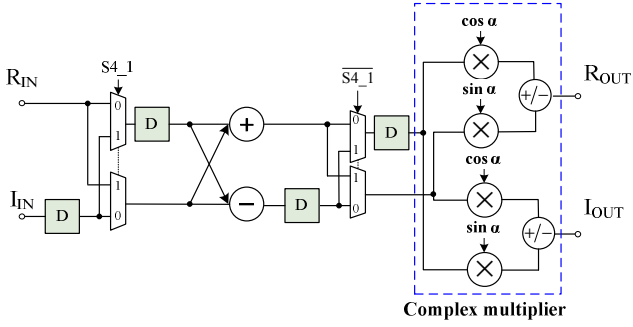

Fig. 6. PE with general complex rotator (PE_TW).

Fig. 5 shows the implementation of the PE for the stages where a $W_{16}$ rotation is required. By using the symmetry property around the unit circle, the rotator only has 3 SAS. Therefore, it only needs to rotate by $W_{16}^0, W_{16}^1$ and $W_{16}^2$, whereas the rest of rotations are obtained by symmetries. If we consider that $W_{16}^1 = (473-196j)/512$, and $W_{16}^2 = (362-362j)/512$, the rotator only needs to multiply its inputs by 473/512, 362/512, or 196/512. This allows for sharing the resources to calculate the multiplications by all these values. With proper setting of control signals, W16_MUL can calculate the multiplication by 473, 362, or 196 using only three real adders.

In the final stage of the radix-$2^k$ flow graphs in Fig. 2, rotations occur both at the upper and lower output paths of each butterfly. Therefore, a complex multiplication is required in these stages. Fig. 6 shows the corresponding PE.

### B. Multi-path SC FFT Architecture

Fig. 7 shows the proposed 128-point 4-parallel MSC FFT architecture. The FFT is realized with a combination of radix-$2^3$ and radix-$2^4$ algorithms, as described in Section III-A. This results in the fact that general complex rotators (PE_TW) are only needed at stage 3 of the architecture. The rest of stages use the simpler PEs described in Section III-A.

TABLE I.  ROTATOR DISTRIBUTION AT STAGES 4 TO 6 OF THE PROPOSED MSC FFT

|  | Stage 4 | Stage 5 | Stage 6 |
|---|---|---|---|
| Path 1 | $W_{16}^0$, $W_{16}^4$ | $W_8^0$ | $W_4^0$ |
| Path 2 | $W_{16}^1$, $W_{16}^5$ | $W_8^1$ | $W_4^0$ |
| Path 3 | $W_{16}^2$, $W_{16}^6$ | $W_8^2$ | $W_4^0$ |
| Path 4 | $W_{16}^3$, $W_{16}^7$ | $W_8^3$ | $W_4^1$ |

In addition to combining radix-$2^3$ and radix-$2^4$, the proposed SC FFT applies the ideas of rotator allocation [17] to simplify the rotators even further. Table I shows the rotations at the parallel paths of stages 4, 5 and 6. One can observe that in the upper path of stage 4, it only has to calculate rotations by $W_{16}^0$ and $W_8^4$, which are trivial. Hence, a PE_W4 is enough here, which much simplifies the rotator complexity. Likewise, other rotators at stages 4, 5 and 6 are simplified by following the same ideas. These rotators correspond to the shaded blue modules in Fig. 7.

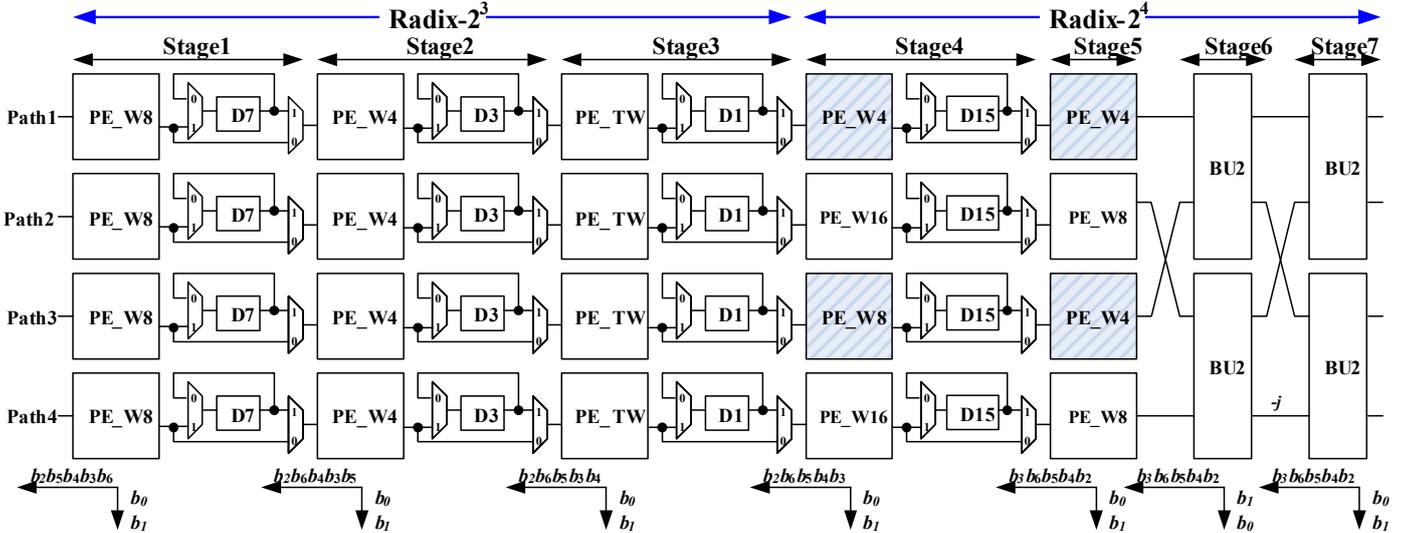

Fig. 7. Proposed 4-parallel radix-$2^3/2^4$ 128-point MSC FFT architecture.

## IV. EXPERIMENTAL RESULTS AND COMPARISON

To verify its performance, the proposed MSC FFT architecture has been synthesized by using the TSMC 90 nm technology. Ten thousand random input patterns of 1 or -1 are applied to the FFT architecture. The experimental results are shown in the second column of Table II. The architecture works at a frequency of 250 MHz, leading to a throughput of 1 GS/s due to the 4 parallel paths. The word length used in the architecture is 12 bits. This results in a SQNR of 42.7 dB. The pre-layout gate-level synthesis results show that the whole FFT architecture area is 0.167 mm$^2$. Finally, the power consumption is 14.81 mW at the clock rate of 250 MHz.

Columns 3 to 7 in Table II show the synthesis results of compared 128-point multi-path FFT architectures. All the compared designs are based either on an MDC or MDF FFT architectures which are derived from their single-path predecessors, namely, SDC and SDF. In contrast, the proposed MSC FFT architecture is the first parallel FFT architecture derived and optimized from the highly-efficient single-path SC FFT.

Together with the proposed architecture, the works of [5], [8] and [9] are all 4-parallel architectures. For a proper area comparison, we have normalized the area figures of all the compared designs according to the following commonly

TABLE II. COMPARISON OF 128-POINT MULTI-PATH FFT ARQUITECTURES

| | This work | [5] | [8] | [9] | [10] | [15] |
|---|---|---|---|---|---|---|
| FFT architecture | 4-parallel MSC | 4-parallel MDF | 4-parallel MRMDF | 4-parallel MDF | 8-parallel MDF | 2-parallel MDC |
| FFT algorithm | radix-$2^3/2^4$ | radix-$2^4$ | radix-8/2 | radix-$2^4$ | radix-$2^3/2^4$ | radix-2 |
| FFT size | 128 | 128 | 128/64 | 128 | 128 | 128 |
| Tech. (nm) | 90 @1.0V | 180 @1.8V | 180 @1.8V | 180 @1.8V | 90 @1V | 65 @0.7V |
| Word length | 12 bits | 12 bits | 10 bits | 10 bits | 8 bits | 12bits |
| Working frequency | 250 MHz | 200 MHz | 250 MHz | 450 MHz | 260 MHz | 160 MHz |
| $A$ (mm$^2$) | 0.167 | - | 3.097 (excl. test module) | - | 0.53 (excl. test module) | 0.34 |
| Gate count | 59049 | 93200 | - | 130000 | | - |
| $P$ (mW) | 14.81 @1 GS/s | 132 @800 MS/s | 175 @1 GS/s | - | 6.8 @409.6 MS/s | 2.12 @317.25 MS/s |
| Normalized Area: $A_N$ (mm$^2$) | 0.167 (26%) | - | 0.774 (100%) | - | 0.53 (68%) | 0.652 (84%) |
| Normalized Power: $P_N$ (mW) | 6.75 @440 MS/s (56%) | 11.2* @440 MS/s (94%) | 11.97 @440 MS/s (100%) | - | 7.3 @440 MS/s (61%) | 8.31 @440 MS/s (69%) |
| SQNR | 42.7dB | 40dB | - | 33dB | 26.4dB | - |
| Throughput rate ($R$: clock rate) | 4$R$ | 4$R$ | 4$R$ | 4$R$ | 8$R$ | 2$R$ |
| Experimental results | Post-synthesis | Post-synthesis | Post-implementation | Post-synthesis | Post-implementation | Post-synthesis |

*: This value has been normalized by multiplying the working frequency rate.

adopted formula, where the proposed design is used as the benchmark design:

$$A_N = \frac{A}{(Tech./90nm)^2}, \quad (2)$$

where $A_N$ is the normalized area, $A$ is the area, and *Tech.* is process feature size of the compared design. For the proposed approach, the process size is $90nm$.

It is observed that the proposed design requires significantly less area than the compared 4-parallel designs: Compared to [5], the area of the proposed design is 61% smaller in terms of gate counts. With respect to [8], the proposed design requires around one-fourth area of [8], while compared to [9], the proposed design consumes less than half of its gate count.

To compare the power consumption, similarly the power data are normalized with respect to the proposed design by using the following formula:

$$P_N = \frac{P}{(Tech./90nm) \times (V/1.0)^2}, \quad (3)$$

where the number 1.0 in the formula is the operating voltage of the proposed implementation, and $V$, $P$ and $P_N$ are the operating voltage, power consumption and normalized power consumption of the compared design, respectively.

At 440 MS/s, the proposed approach consumes only 60% and 56% of the total power consumption in [5] and [8], respectively. This represents savings of 40% or larger in power consumption.

Finally, the SQNR is calculated as explained in [20]. By comparing the results in the table, it can be observed that the proposed approach achieves the highest SQNR among all the 128-point multi-path FFT designs.

V. CONCLUSIONS

In this work, we have presented the first multi-path SC FFT (MSC FFT). The demonstration design is a 128-point 4-parallel radix-$2^3/2^4$ MSC FFT. The use of an SC-based processing element facilitates the minimization of the hardware resources in the architecture. Furthermore, the utilization of radix-$2^3/2^4$ FFT combined with optimized rotator allocation results in highly hardware-efficient rotator architectures. Compared with the existing designs with the same FFT size and degree of parallelism, the proposed architecture achieves savings by more than 40% in both area and power consumption, while provide very high throughput. The proposed architecture can be extended to higher parallelism than four with much higher throughputs which can meet the requirements of current and future most demanding communication systems (with throughput of more than 10Gpbs), such as 5G mobile systems, IEEE802.11ax and IEEE 802.11ay systems. Those application designs are currently under investigation.